\documentclass[sigconf]{acmart}

\usepackage{todonotes}
\usepackage{csquotes}
\usepackage{hyperref}
\usepackage{comment}
\usepackage{xcolor}
\usepackage{multirow}
\usepackage{colortbl}
\usepackage{subfig}
\usepackage{tipa}
\usepackage{amsmath}
\usepackage{calc}

\definecolor{highlight1}{RGB}{223, 156, 0} 
\definecolor{highlight2}{RGB}{191, 61, 58}  

\definecolor{Benchmark_Other_Dataset}{RGB}{198, 224, 180}
\definecolor{Other_Use_Cases}{RGB}{172, 194, 219}
\definecolor{Engagement_With_NGO}{RGB}{255, 229, 180}
\definecolor{Protection_Other_AA}{RGB}{255, 171, 185}
\definecolor{IMDB_individuals}{RGB}{171, 171, 171}

\AtBeginDocument{%
  \providecommand\BibTeX{{%
    \normalfont B\kern-0.5em{\scshape i\kern-0.25em b}\kern-0.8em\TeX}}}

\setcopyright{cc}
\setcctype[4.0]{by-nc-nd}

\copyrightyear{2024}
\acmYear{2024}
\acmConference[]{}{}{}
\acmBooktitle{}
\acmDOI{10.1145/3630106.3658936}
\acmISBN{979-8-4007-0450-5/24/06}

\begin{document}



\title[A Semi-automated Text Sanitization Tool for Mitigating the Risk of Whistleblower Re-Identification]{Silencing the Risk, Not the Whistle: A Semi-automated Text Sanitization Tool for Mitigating the Risk of Whistleblower Re-Identification}

\newcommand{\toolname}{our tool}



\author{Dimitri Staufer}
\affiliation{%
  \institution{Technische Universität Berlin}
  \city{Berlin}
  \country{Germany}}
\email{staufer@tu-berlin.de}

\author{Frank Pallas}
\affiliation{%
  \institution{Paris Lodron University Salzburg}
  \city{Salzburg}
  \country{Austria}}
\email{frank.pallas@plus.ac.at}

\author{Bettina Berendt}
\affiliation{%
  \institution{TU Berlin, Weizenbaum Institute, and KU Leuven}
  \city{Berlin and Leuven}
  \country{Germany and Belgium}}
\email{berendt@tu-berlin.de}

\renewcommand{\shortauthors}{Staufer, et al.}

\begin{abstract}
Whistleblowing is essential for ensuring transparency and accountability in both public and private sectors. However, (potential) whistleblowers often fear or face retaliation, even when reporting anonymously. The specific content of their disclosures and their distinct writing style may re-identify them as the source. Legal measures, such as the EU Whistleblower Directive, are limited in their scope and effectiveness. Therefore, computational methods to prevent re-identification are important complementary tools for encouraging whistleblowers to come forward. However, current text sanitization tools follow a one-size-fits-all approach and take an overly limited view of anonymity. They aim to mitigate identification risk by replacing typical high-risk words (such as person names and other labels of named entities) and combinations thereof with placeholders. Such an approach, however, is inadequate for the whistleblowing scenario since it neglects further re-identification potential in textual features, including the whistleblower's writing style. Therefore, we propose, implement, and evaluate a novel classification and mitigation strategy for rewriting texts that involves the whistleblower in the assessment of the risk and utility. Our prototypical tool semi-automatically evaluates risk at the word/term level and applies risk-adapted anonymization techniques to produce a grammatically disjointed yet appropriately sanitized text. We then use a Large Language Model (LLM) that we fine-tuned for paraphrasing to render this text coherent and style-neutral. We 
evaluate \toolname's effectiveness
using
court cases from the European Court of Human Rights (ECHR) and excerpts from a real-world whistleblower testimony and measure the protection against authorship attribution attacks and utility
loss statistically using the popular IMDb62 movie reviews dataset, which consists of 62 individuals. Our method can significantly reduce authorship attribution accuracy from 98.81\% to 31.22\%, while preserving up to 73.1\% of the original content's semantics, as measured by the established cosine similarity of sentence embeddings.
\end{abstract}

\ccsdesc[500]{Security and privacy~Privacy protections}
\ccsdesc[500]{Security and privacy~Pseudonymity, anonymity and untraceability}
\ccsdesc[500]{Computing methodologies~Natural language processing}
\ccsdesc[500]{Information systems~Retrieval tasks and goals}
\ccsdesc[500]{Security and privacy~Human and societal aspects of security and privacy}
\ccsdesc[500]{Human-centered computing~Collaborative and social computing systems and tools}
\ccsdesc[300]{Applied computing~Law, social and behavioral sciences}
\ccsdesc[300]{Information systems~Web applications}
\ccsdesc[100]{Information systems~Data mining}

\keywords{Text Sanitization, Whistleblower Anonymity, Authorship Obfuscation, Fine-tuning Language Models, LLM-based Rephrasing}




\maketitle

\section{Introduction}

In recent years, whistleblowers have become ``a powerful force'' for transparency and accountability, not just in the field of AI \cite{ainow2019}, but also in other technological domains and across both private- and public-sector organizations. Institutions such as the AI Now Institute \cite{ainow2019} or the IEEE Global Initiative on Ethics of Autonomous and Intelligent Systems \cite{IEEE2019} have emphasized the key role of whistleblower protection for societal well-being and often also the organizations' own interests \cite{HauserEtAl2019}. 
However, whistleblowing may be a threat for the organizations whose malfeasance is being revealed; thus (potential) whistleblowers often fear or face retaliation. Computationally-supported anonymous reporting seems to be a way forward, but even if reporting frameworks are sufficiently secure system- and network-wise, the report itself may allow inferences towards the whistleblower's identity due to its content and the whistleblower's writing style. 
Non-partisan organizations such as Whistleblower-Netzwerk e.V. (WBN) provide guidance on concise writing. Our interactions with WBN confirm that whistleblower testimonies often include unnecessary personal details.

Existing approaches modifying the texts of such reports appear promising, but they take an overly limited view of anonymity and -- like whistleblower protection laws -- address only parts of the problem. This is detailed in Section \ref{sec_relwork}.
To improve on these approaches, we propose, implement, and evaluate a novel classification and mitigation strategy for rewriting texts that puts the whistleblower into the loop of assessing risk and utility.

Our contributions are threefold.
First (Section \ref{sec_concepts}), we analyse the interleaved contributions of different types of identifiers in texts to derive a description of the problem for anonymous whistleblowing in terms of a trade-off between risk (identifiability of the whistleblower) and utility (of the rewritten text retaining sufficient information on the specific event details). 
We derive a strategy for assigning re-identification risk levels of concern to textual features composed of an automated mapping and an interactive adjustment of concern levels. 
Second (Section \ref{sec:implementation}), we describe \toolname which implements this strategy. It applies (i) the word/term-to-concern mapping using natural language processing to produce a sanitized but possibly ungrammatical intermediate text version, (ii) a Large Language Model (LLM) that we fine-tuned for paraphrasing to render this text coherent and style-neutral, and (iii) interactivity to draw on the user's context knowledge. 
Third (Section \ref{sec_eval}), we evaluate the resulting
risk-utility trade-off. We measure the protection against authorship attribution attacks and utility loss statistically using an established
benchmark dataset and show that it can significantly reduce authorship attribution accuracy while retaining utility. We also evaluate our \toolname's effectiveness in masking direct and quasi-identifiers using the Text Anonymization Benchmark \cite{pilan2022text} and demonstrate its effectiveness on excerpts from a real-world whistleblower testimony. Section \ref{sec_summary} sketches current limitations and future work. Section \ref{sec_ethics} describes ethical considerations and researchers' positionality, and it discusses possible adverse impacts.

\section{Background and Related Work}
\label{sec_relwork}

This section describes the importance of, and threats to, whistleblowing (Section \ref{sec_challenges}) and 
the promises and conceptual and practical challenges of ``anonymity'' in reporting  (Section \ref{sec_anonymity}). We  survey related work on the anonymization/de-identification of text and argue why it falls short in supporting whistleblowing (Section \ref{relatedWork}).

\subsection{Challenges of Safeguarding Whistleblowers}
\label{sec_challenges}

Whistleblowers play a crucial role in exposing wrongdoings like injustice, corruption, and discrimination in organizations \cite{near1985organizational,bosua2014going}. However, their courageous acts often lead to negative consequences, such as subtle harassment and rumors, job loss and blacklisting, and, in extreme cases, even death threats \cite{martin2003illusions, sawyer2010necessary, mcglynn2014private}. In Western nations, whistleblowing is largely viewed as beneficial to society \cite{van2022whistleblowing}, leading to protective laws like the US Sarbanes-Oxley Act of 2002 and the European Union's \enquote{Whistleblowing Directive} (Directive 2019/1937). The latter, for example, mandates the establishment of safe reporting channels and protection against retaliation. It also requires EU member states to provide whistleblowers with legal, financial, and psychological support. However, the directive faces criticism for its limitations. Notably, it does not cover all public-sector entities \cite[p. 3]{terracol2019building} and leaves key decisions to member states' discretion \cite[p. 652]{abazi2020european}. This discretion extends to the absence of mandatory anonymous reporting channels and permits states to disregard cases they consider \enquote{clearly minor}, leaving whistleblowers without comprehensive protection for non-material harms like workplace bullying \cite[p. 3]{terracol2019building}. Furthermore, according to \citet{white2018matter}, the directive's sectoral approach and reliance on a list of specific EU laws causes a patchwork of provisions, creating a complex and possibly confusing legal environment, particularly for those sectors impacting human rights and life-and-death situations.

Last but not least, organizations often react negatively to whistleblowing due to the stigma of errors, even though  recognizing these mistakes would be key to building a culture of responsibility \cite[p. 12]{berendt2022whistleblower} and improving organizations and society \cite{weingardt2004fehler}. The reality for whistleblowers is thus fraught with challenges, from navigating legal uncertainties to dealing with public perception \cite{sachdeva2022exploring, leite2021whistleblowing, saade2023women}, leaving many whistleblowers with no option but to report their findings anonymously \cite{rothschild1999whistle}. However, ``anonymous'' reporting channels alone do not guarantee anonymity \cite{berendt2022whistleblower}.

\subsection{Anonymity, (De-)anonymization, and (De-/Re-)Identification}
\label{sec_anonymity}

Anonymity is not an alternative between being identified uniquely or not at all, but ``the state of being not identifiable within a set of subjects [with potentially the same attributes], the anonymity set'' \cite[p.9]{PfitzmannHansen2010}. Of the manifold possible approaches towards this goal, state-of-the-art whistleblowing-support software as well as legal protections (where existing) focus on
{\em anonymous communications} \cite{berendt2022whistleblower}. This, however, does not guarantee {\em anonymous reports}. Instead, a whistleblower's anonymity may still be at risk due to several factors, including: (i) surveillance technology, such as browser cookies, security mechanisms otherwise useful to prevent unauthenticated uses, cameras, or access logs, (ii) the author's unique writing style, and (iii) the specific content of the message \cite{marcum2019blowing}. \citet{berendt2022whistleblower} refer to the latter as \enquote{epistemic non-anonymizability}, 
i.e., the risk of being identified based on the unique information in a report, particularly when the information is known to only a few individuals. In some cases, this may identify the whistleblower uniquely. 

\label{identifier_classes}

Terms and their understanding in the domain of anonymity vary. We use the following nomenclature:
{\em anonymization} is a modification of data that increases the size of the anonymity set of the person (or other entity) of interest; conversely, {\em de-anonymization} decreases it (to some number $k \geq 1)$. De-anonymization to $k=1$, which includes the provision of an identifier (e.g., a proper name), is called  {\em re-identification}. The removal of some identifying information (e.g., proper names), called {\em de-identification}, often but not necessarily leads to anonymization \cite{ben2022anonymization, weggenmann2018syntf}.

In structured data, direct identifiers (e.g., names or social security numbers) are unique to an individual, whereas quasi-identifiers like age, gender, or zip code, though not unique on their own, can be combined to form unique patterns. Established mathematical frameworks for quantifying anonymity, such as Differential Privacy (DP) \cite{dwork2006differential}, and metrics such as k-anonymity \cite{sweeney2002k}, along with their refinements \cite{machanavajjhala2007diversity, li2006t}, can be used when anonymizing datasets.

Unstructured data such as text, which constitutes a vast majority of the world's data, requires its own safeguarding methods, which fall into two broader categories \cite{lison2021anonymisation}. The first, NLP-based text sanitization, focuses on linguistic patterns to reduce (re-)identification risk. The second, privacy-preserving data publishing (PPDP), involves methods like noise addition or generalization to comply with pre-defined privacy requirements \cite{domingo2016database}.

\subsection{Related Work: Text De-Identification and Anonymization, Privacy Models, and Adversarial Stylometry} \label{relatedWork}

De-identification methods in text sanitization mask identifiers,
primarily using named entity recognition (NER) techniques. These methods, largely domain-specific, have been particularly influential in clinical data de-identification, as evidenced, for instance, by the 2014 i2b2/UTHealth shared task \cite{stubbs2015automated}. However, they do not or only partially address the risk of indirect re-identification \cite{mehta2019towards, ben2022anonymization}. For example, \citet{sanchez2012detecting, sanchez2013automatic, sanchez2014utility} make the simplifying assumption that replacing noun phrases which are rare in domain-specific corpora or on the web with more general ones offers sufficient protection. Others use recurrent neural networks \cite{dernoncourt2017identification, liu2017identification}, reinforcement learning \cite{xu2019privacy}, support vector machines \cite{uzuner2008identifier}, or pre-trained language models \cite{johnson2020deidentification} to identify and remove entities that fall into pre-defined categories. However, all of these approaches ignore or significantly underestimate the actual risks of context-based re-identification.

More advanced anonymization methods, in turn, also aim to detect and remove identifiers that do not fit into the usual categories of named entities or are hidden within context. For example, \citet{reddy2016obfuscating} detect and obfuscate gender, and \citet{adams2019anonymate} introduce a human-annotated multilingual corpus containing 24 entity types and a pipeline consisting of NER and co-reference resolution to mask these entities. In a more nuanced approach, \citet{papadopoulou2022neural} developed a \enquote{privacy-enhanced entity recognizer} that identifies 240 Wikidata properties linked to personal identification. Their approach includes three key measures to evaluate if a noun phrase needs to be masked or replaced by a more general one \cite{olstad2023generation}. The first measure uses RoBERTa \cite{liu2019roberta} to assess how \enquote{surprising} an entity is in its context, assuming that more unique entities carry higher privacy risks. The second measure checks if web search results for entity combinations mention the individual in question, indicating potential re-identification risk. Lastly, they use a classifier trained with the Text Anonymization Benchmark (TAB) corpus \cite{pilan2022text} to predict masking needs based on human annotations.

\citeauthor{kleinberg2022textwash}'s \cite{kleinberg2022textwash} \enquote{Textwash} employs the BERT model, fine-tuned on a dataset of 3717 articles from the British National Corpus, Enron emails, and Wikipedia. The dataset was annotated with
entity tags such as \enquote{PERSON\_FIRSTNAME}, \enquote{LOCATION}, and an \enquote{OTHER\_IDENTIFYING\_ATTRIBUTE} category for indirect re-identification risks, along with a \enquote{NONE} category for tokens that are non-re-identifying. A quantitative evaluation (0.93 F1 score for detection accuracy, minimal utility loss in sentiment analysis, and part-of-speech tagging) and its qualitative assessment (82\% / 98\%   success in anonymizing famous / semi-famous individuals) show promise. However, the more recent gpt-3.5-turbo can re-identify 72.6\% of the celebrities from Textwash's qualitative study on the first attempt, highlighting the evolving complexity of mitigating the risk of re-identification in texts \cite{patsakis2023man}.

In PPDP, several privacy models for structured data have been adapted for privacy guarantees in text. While most are theoretical \cite{lison2021anonymisation}, \enquote{C-sanitise} \cite{sanchez2016c} determines the disclosure risk of a certain term \textit{t} on a set of entities to protect (\textit{C}), given background knowledge \textit{K}, which by default is the probability of an entity co-occurring with a term \textit{t} in the web. Additionally, DP techniques have been adapted to text, either for generating synthetic texts \cite{fernandes2019generalised} or for obscuring authorship in text documents \cite{weggenmann2018syntf}. This involves converting text into word embeddings, altering these vectors with DP techniques, and then realigning them to the nearest words in the embedding model \cite{yue2021differential,zhao2022survey}. However, \enquote{word-level differential privacy} \cite{mattern2022limits} faces challenges: it maintains the original sentence length, limiting variation, and can cause grammatical errors, such as replacing nouns with unrelated adjectives, due to not considering word types.

Authorship attribution (AA) systems use stylistic features such as vocabulary, syntax, and grammar to identify an author. State-of-the-art approaches involve using Support Vector Machines \cite{yadav2020authorship, tyo2022state}, and more recently, fine-tuned LLMs like BertAA \cite{fabien2020bertaa, altakrori2021topic, tyo2022state}. The \enquote{Valla} benchmark and software package
standardizes evaluation methods and includes fifteen diverse datasets \cite{tyo2022state}. Contrasting this, adversarial stylometry modifies an author's writing style to reduce AA systems' effectiveness \cite{stuart2013identifying}. Advancements in machine translation \cite{wang2021progress} have also introduced new methods based on adversarial training \cite{shetty2018a4nt}, though they sometimes struggle with preserving the original text's meaning. Semi-automated tools, such as \enquote{Anonymouth} \cite{mcdonald2012use}, propose modifications for anonymity in a user's writing, requiring a significant corpus of the user's own texts. Moreover, recent advances in automatic paraphrasing using fine-tuned LLMs demonstrated a notable reduction in authorship attribution, but primarily for shorter texts \cite{mattern2022limits}.

To the best of our knowledge, there is no -- and maybe there can be no -- complete list of textual features contributing to the re-identification of individuals in text.
As \citet{narayanan2010myths} highlight, \enquote{any attribute can be identifying in combination with others} [p. 3]. In text, we encounter elements like characters, words, and phrases, each carrying varying levels of meaning \cite{feldman2007text}. Single words convey explicit lexical meaning as defined by a vocabulary (e.g. \enquote{employee}), while multiple words are bound by syntactic rules to express more complex thoughts implicitly in phrases (\enquote{youngest employee}) and sentences (\enquote{She is the youngest employee}). 

In addition, the \citet{edps_aepd_2021} state that anonymization can never be fully automated and needs to be \enquote{tailored to the nature, scope, context and purposes of processing as well as the risks of varying likelihood and severity for the rights and freedoms of natural persons} [p. 7]. 

To take these insights and limitations into account, our semi-automated text sanitization tool leverages insights on the removal of identifying information but involves the whistleblower (the user) in the decision-making process.

\section{Risk modelling and risk mitigation approach}
\label{sec_concepts}

In this section, we derive the problem statement (Section \ref{sec_problem_statement}) from an analysis of different identifier types (Section \ref{sec_identifier_types}). Following an overview of our approach (Section \ref{sec_mitigation_overview}), we detail the anonymization operations for textual features (Section \ref{sec_an_ops}) and the automatic assignment of default concern levels (Section \ref{automatic_LvC_estimation}).

\subsection{Identifier Types, Author Identifiability, and Event Details in the Whistleblowing Setting}
\label{sec_identifier_types}

Whistleblowing reports convey information about persons, locations, and other entities. At least some of them need to be {\em identified} in order for the report to make any sense. 
The following fictitious example consists of three possible versions of a report in order to illustrate how different types of identifiers may contribute to the re-identification of the anonymously reporting employee Jane Doe, a member of the Colours and Lacquer group in the company COLOURIFICS.

\begin{description}
\item[V1] On 24 January 2023, John Smith poured polyurethane resin into the clover-leaf-shaped sink of room R23.
\item[V2] After our group meeting on the fourth Tuesday of January 2023, the head of the Colours and Lacquer Group poured a toxin into the sink of room R23.
\item[V3] Somebody poured a liquid into a recepticle on some date in a room of the company.
\end{description}

In V1, “John Smith” is the {\em lexical identifier}\footnote{The classification of identifiers is due to Phillips \cite{Phillips04}. Note that all types of identifiers can give rise to {\em personal data.}. in the sense of the EU's General Data Protection Regulation (GDPR), Article 4(1): ``any information which is related to an identified or identifiable natural person'', or {\em personally identifiable data} in the senses used in different US regulations. See 
\cite{DBLP:conf/apf/CostaR20} for legal aspects in the
context of whistleblowing.}
of the COLOURIFICS manager John Smith, as is “24 January 2023” of that date. Like John Smith, room R23 is a unique named entity in the context of the company and also identified lexically. “Polyurethane resin” is the lexical identifiers of a toxin (both are common nouns rather than names of individual instances of their category). The “clover-leaf-shaped” serves as a descriptive identifier of the sink. In V2, John Smith is still identifiable via the {\em descriptive identifier} “head of the Colours and Lacquer Group”, at least on 24 January 2023 (reconstructed with the help of a calendar and COLOURIFIC’s personnel files). “Our” group meeting is an {\em indexical identifier} that signals that the whistleblower is one of the, say five employees in the Colours and Lacquer Group.

The indexical information is explicit in V2 given the background knowledge that only employees in this group were co-present (for example, in the company’s key-card logfiles). The same information may be implicit in V1 (if it can be seen from the company’s organigram who John Smith is and who works in his group).
Both versions provide for the inference that Jane Doe or any of her four colleagues must have been the whistleblower. If, in addition, only Jane Doe stayed behind “after the meeting”, that detail in V2 descriptively identifies her uniquely\footnote{If John Smith knows that only she observed him, she is also uniquely identified in V1, but for the sake of the analysis, we assume that only recorded data/text constitute the available knowledge.}. 
V3 contains only identifiers of very general categories. Many other variants are possible (for example, referencing, in a V4, “the head of our group”, which would enlarge the search space to all groups that had a meeting in R23 that day).

The example illustrates the threats (i)-(iii) of Section \ref{sec_anonymity}. It also shows
that the whistleblower’s ``anonymity'' (or lack thereof) is only one aspect of a more general and graded picture of who and what can be identified directly, indirectly, or not at all – and what this implies for the whistleblower’s safety as well as for the report’s effectiveness. 

Inspired by Domingo-Ferrer’s \cite{domingo2007three} three types of (data) privacy, we distinguish between the identifiability of the whistleblower Jane Doe ({\em author}
\footnote{We assume that the potential whistleblower is also the author of the report. This is the standard setting. Modifications for the situation in which a trusted third party writes the report on their behalf are the subject of future work.}
{\em identifiability $A_{id}$}) and descriptions of the event or other wrongdoing, including other actors ({\em event details $E_{dt}$}). 
Given the stated context knowledge, we obtain an anonymity set of size $k=1$ for John Smith in V1 and V2. Jane Doe is in an anonymity set of size $k=5$ or even $k=1$ in V2. In V1, that set may be of size $k=5$ (if people routinely work only within their group) or larger (if they may also join other groups). Thus, the presence of a name does not necessarily entail a larger risk.
Both are in an anonymity set containing all the company’s employees at the reported date in V3 (assuming no outsiders have access to company premises). The toxin and the sink may be in a smaller anonymity set in V1 than in V2 or V3, and they could increase further (for example, if only certain employees have access to certain substances). Importantly, the identifiability of people and other entities in $E_{dt}$ can increase the identifiability of the whistleblower. 

V3 illustrates a further challenge: the misspelled receptacle may be a typical error of a specific employee, and the incorrect placement of the temporal before the spatial information suggests that the writer may be a German or Dutch native speaker. In addition to errors, also correct variants carry information that stylometry can use for authorship attribution, which obviously can have a large effect on $A_{id}$. 

The whistleblower would, on the one hand, want to reduce all such identifiabilities as much as possible. On the other hand, the extreme generalization of V3 creates a meaningless report that neither the company nor a court would follow up on. This general problem can be framed in terms of risk and utility, which will be described next.

\subsection{The Whistleblowing Text-Writing Problem: Risk, Utility, And  Many Unknowns} 
\label{sec_problem_statement}

A potential whistleblower faces the following problem: “make $A_{id}$ as small as possible while retaining as much $E_{dt}$ as necessary”. We propose to address this problem by examining the text and possibly rewriting it. 

In principle, this is an instance of the oft-claimed trade-off between privacy (or other risk) and utility. In a simple world of known repositories of structured data, one could aim at determining the identifying problem (e.g., by database joins to identify the whistleblower due to some attributive information they reveal about themselves and by multiple joins for dependencies such as managers and teams) and compute how large the resulting anonymity set (or $A_{id}$ as its inverse) is. Given a well-defined measure of information utility, different points on the trade-off curve would then be well-defined and automatically derivable solutions to a mathematical optimization problem. 

However, texts offer a myriad of ways to express a given relational information. The space of information that could be cross-referenced, sometimes in multiple steps, is huge and often unknown to the individual. Consequently, in many cases, it is not possible to determine the anonymity set size with any mathematical certainty. In addition, setting a threshold could be dangerous: even if the anonymity set is $k > 1$, protection is not guaranteed – for example, the whole department of five people could be fired in retaliation. 
At the same time, exactly how specific a re-written text
needs to be about $A_{id}$ and $E_{dt}$ in order to make the report legally viable
\footnote{``a situation in which a plan, contract, or proposal is able to be legally enforced'', \url{https://ludwig.guru/s/legally+viable}, retrieved 2024-01-02} cannot be decided without much more context knowledge. For example, the shape of the sink into which a toxic substance is poured probably makes no difference to the illegality, whereas the identity of the substance may affect it.

These unknowns have repercussions both for tool design (Section \ref{sec_mitigation_overview}) and for evaluation design (Section \ref{sec_evaluation_design}).

\subsection{Risk Mitigation Approach and Tool Design: Overview}
\label{sec_mitigation_overview}

Potential whistleblowers would be ill-served by any fully automated tool that claims to be able to deliver a certain mathematically guaranteed anonymization. Instead, we propose to provide them with a semi-automated tool that does have some “anonymity-enhancing defaults” that illustrate with the concrete material how textual elements can be identifying and how they can be rendered less identifying. Our tool starts with the heuristic default assumption that identifiability is potentially {\em always} problematic and then lets the user steer \toolname\ by specifying how “concerning” {\em specific} individual elements are and choosing, interactively, the treatment of each of them that appears to give the best combination of $A_{id}$ and $E_{dt}$. By letting the author/user assign these final risk scores in the situated context of the evolving text, we enable them to draw on a maximum of implicit context knowledge.

Our approach and tool proceed through several steps. We first determine typical textual elements that can constitute or be part of the different types of identifiers. As can be seen in Table \ref{table_overview}, most of them can affect $A_{id}$ and $E_{dt}$.

Since identification by name (or, by extension, pronouns that co-reference names) does not even need additional background knowledge and since individuals are more at risk than generics, we classify some textual features as “highly concerning”, others as having “medium concern”, and the remainder as “potentially concerning”. 
We differentiate between two types of proper nouns. Some names refer to typical “named entities”, which include, in particular, specific people, places, and organizations, as well as individual dates and currency amounts. These pose particular person-identification risk in whistleblowing scenarios.\footnote{PERSON, GPE (region), LOC (location), EVENT, LAW, LANGUAGE, DATE, TIME, PERCENT, MONEY, QUANTITY, and ORDINAL}
``Other proper nouns'', such as titles of music pieces, books and artworks generally only pose medium risk. For stylometric features, we explicitly categorize out-of-vocabulary words, misspelled words, and words that are surprising given the overall topic of the text. Other low-level stylometric features, such as punctuation patterns, average word and sentence length, or word and phrase repetition, are not (and in many cases, such as with character n-gram pattern, cannot be \cite{lagutina2019survey}) explicitly identified. Instead, we implicitly/indirectly account for them as a byproduct of the LLM-based rephrasing. For all other parts of speech, we propose to use replacement strategies based on data-anonymization operations that are proportional to the risk (Table \ref{table_strategies}). Given the complexities of natural language and potential context information, the latter two operations are necessarily heuristic; thus, \toolname\ applies the classification and the risk mitigation strategy as a default which can then be adapted by the user.

\begin{table}[H]
\centering
\caption{Overview of the approach from identifier types to default risk.}
\Description{Overview of the approach from identifier types to default risk.}
\label{tab:identifier_risk}
\begin{tabular}{@{}p{1.8cm}llp{1cm}@{}}
\toprule
\textbf{Identifier Type} & \textbf{Textual Feature} &
$\boldsymbol{\mathit{A}}_{\boldsymbol{\mathit{id}}}/\boldsymbol{\mathit{E}}_{\boldsymbol{\mathit{dt}}}$ & 
\textbf{Default Risk} \\ \midrule
Lexical & Names of named entities & $A_{id}$,$E_{dt}$ & High \\
Lexical & Other proper nouns & $E_{dt}$ & Medium \\
Indexical & Pronouns & $A_{id}$,$E_{dt}$ & High \\
Descriptive & Common nouns & $E_{dt}$,($A_{id}$) & Potential \\
Descriptive & Modifiers & $E_{dt}$,($A_{id}$) & Potential \\
Descriptive\newline(via pragmatic inferences) & Out-of-vocabulary words\textsuperscript{\hyperlink{fn:footnote1}{a}} & $A_{id}$, ($E_{dt}$) & Medium \\
& Misspelled words\textsuperscript{\hyperlink{fn:footnote1}{a}} & $A_{id}$ & Medium \\
& Surprising words\textsuperscript{\hyperlink{fn:footnote2}{b}} & $A_{id}$ & Medium \\
& Other stylometric features & $A_{id}$ & N/A\textsuperscript{\hyperlink{fn:footnote3}{c}} \\
\bottomrule
\end{tabular}
\label{table_overview}
\end{table}
{\raggedright 
\textsuperscript{\hypertarget{fn:footnote1}{a}}Treated as noun. 
\textsuperscript{\hypertarget{fn:footnote2}{b}}Nouns or proper nouns.
\textsuperscript{\hypertarget{fn:footnote3}{c}}Not explicitly specified. Indirectly accounted for through rephrasing.}

\definecolor{potentialconcern}{RGB}{247, 247, 247}
\definecolor{mediumconcern}{RGB}{255, 226, 153}
\definecolor{highconcern}{RGB}{255, 171, 157}

\begin{table}[H]
\centering
\caption{Mitigation strategies based on assigned risk (LvC = level of concern, NaNEs = names of named entities, OPNs = other proper nouns, CNs = common nouns, Mods = modifiers, PNs = proper nouns, OSFs = other stylometric features).}
\Description{Mitigation strategies based on assigned risk.}
\label{tab:risk_mitigation}
\begin{tabular}{@{}lcccccc@{}}
\toprule
\textbf{LvC} & \textbf{NaNEs} & \textbf{OPNs} & \textbf{CNs} & \textbf{Mods} & \textbf{PNs} &
\textbf{OSFs} \\ 
\midrule
\cellcolor{highconcern}\textbf{High} & Suppr. & Suppr. & Suppr. & Suppr. & Suppr. &
Pert. \\
\cellcolor{mediumconcern}\textbf{Medium} & Pert. & Generl. & Generl. & Pert. & Suppr. &
Pert. \\
\bottomrule
\end{tabular}
\label{table_strategies}
\end{table}

\definecolor{potentialconcerndarker}{RGB}{247, 247, 247}
\definecolor{mediumconcerndarker}{RGB}{162, 104, 0}
\definecolor{highconcerndarker}{RGB}{191, 48, 35}
\colorlet{mixedconcerncolor}{mediumconcerndarker!50!highconcerndarker}

\subsection{Anonymization Operations for Words and Phrases}

\label{sec_an_ops}

In our sanitization pipeline, we conduct various token removal and replacement operations based on each token's \textit{POS tag} and its assigned level of concern (LvC), which can be \enquote{potentially concerning}, \enquote{medium concerning}, or \enquote{highly concerning}. Initially, we consider all common nouns, proper nouns, adjectives, adverbs, pronouns, and named entities\footnote{By this, we mean names of named entities, e.g. \enquote{Berlin} for GPE, but we use \texttt{named entities} instead for consistency with other literature.} as potentially concerning. Should the user or our automatic LvC estimation (see subsection \ref{automatic_LvC_estimation}) elevate the concern to either medium or high, we apply anonymization operations that are categorized into generalization, perturbation, and suppression. Specific implementation details are elaborated on in section \ref{sec:implementation}.

\subsubsection{Generalization}
The least severe type of operation targets \texttt{common nouns} and \texttt{other proper nouns} marked as \textcolor{mediumconcerndarker}{\textbf{medium concerning}}. We assume their specificity (not necessarily their general meaning) poses re-identification risks. Thus, more general terms can be used to preserve meaning while mitigating the risk of re-identification.

\begin{itemize}
    \item \texttt{Common nouns} like \enquote{car} are replaced with hypernyms from WordNet, such as \enquote{vehicle}.
    \item \texttt{Other proper nouns} become broader Wikidata terms, e.g. \enquote{political slogan} for \enquote{Make America Great Again}.
\end{itemize}

\subsubsection{Perturbation}
This applies to \texttt{modifiers}%
\footnote{The current version of \toolname\ considers only adjectives and adverbs as modifiers.}
and \texttt{named entities} annotated as \textcolor{mediumconcerndarker}{\textbf{medium concerning}}. In this process, original words are retained but are assigned zero weight in the paraphrase generation, along with their synonyms and inflections. This approach relies on the LLM to either (a) find similar but non-synonymous replacement words or (b) completely rephrase the sentence to exclude these words. For example, \enquote{\textit{Microsoft}, the \textit{giant} tech company, ...} could be paraphrased as \enquote{A leading corporation in the technology sector...}.

\subsubsection{Suppression}
The most severe type of operation is applied to \texttt{common nouns}, \texttt{other proper nouns}, \texttt{modifiers} and \texttt{named entities} annotated as \textcolor{highconcerndarker}{\textbf{highly concerning}}, and to \texttt{pronouns} that are either \textcolor{mediumconcerndarker}{\textbf{medium concerning}} or \textcolor{highconcerndarker}{\textbf{highly concerning}}. We assume these words are either too unique or cannot be generalized.

\begin{itemize}
    \item For \texttt{common nouns} and \texttt{other proper nouns}, dependent phrases are omitted (e.g., \enquote{We traveled \textit{to the London Bridge} in a bus.} becomes \enquote{We traveled in a bus.}).
    \item \texttt{Modifiers} are removed (e.g., \enquote{He used to be the \textit{principal} dancer} becomes \enquote{He used to be a dancer}).
    \item \texttt{Named entities} are replaced with nondescript phrases (e.g., \enquote{Barack Obama} becomes \enquote{certain person}).
    \item \texttt{Pronouns} are replaced with \enquote{somebody} (e.g., \enquote{\textit{He} drove the bus.} becomes \enquote{Somebody drove the bus.}).
\end{itemize}

\subsection{Automatic Level of Concern (LvC) Estimation} \label{automatic_LvC_estimation}

In our whistleblowing context, we deem the detection of outside-document LvC via search engine queries, as proposed by \citet{papadopoulou2022neural} (refer to related work in \ref{relatedWork}), impractical. This is because whistleblowers are typically not well-known, and the information they disclose is often novel, not commonly found on the internet. Therefore, instead of relying on external data, we focus on inner-document LvC, setting up a rule-based system and allowing users to adjust the LvC based on their contextual knowledge. Further, we assume that this pre-annotation of default concern levels raises awareness for potential sources of re-identification.

\begin{itemize}
\item \texttt{Common nouns} and \texttt{modifiers}, by default, are \textbf{potentially concerning}. As fundamental elements in constructing a text's semantic understanding, they could inadvertently reveal re-identifying details like profession or location. However, without additional context, their LvC is not definitive.
\item \texttt{Other proper nouns}, \texttt{unexpected words}, \texttt{misspelled words} and \texttt{out-of-vocabulary words} default to \textcolor{mediumconcerndarker}{\textbf{medium concerning}}. Unlike categorized named entities, \texttt{other proper nouns} only indirectly link to individuals, places, or organizations. Unexpected words may diminish anonymity, according to \citet{papadopoulou2022neural}, while misspelled or out-of-vocabulary words can be strong stylometric indicators.
\item \texttt{Named entities} are considered \textcolor{highconcerndarker}{\textbf{highly concerning}} by default, as they directly refer to specific entities in the world, like people, organizations, or locations, posing a significant re-identification risk.
\end{itemize}

\section{Implementation} \label{sec:implementation}

Our semi-automated text sanitization tool consists of a sanitization pipeline (Sections \ref{sec_impl_1} and \ref{sec_impl_2})
and a user interface
(Section \ref{sec_impl_3}). The pipeline uses off-the-shelf Python NLP libraries (\textit{spaCy}, \textit{nltk}, \textit{lemminflect}, \textit{constituent\_treelib}, \textit{sentence-transformers}) and our paraphrasing-tuned FLAN T5 language model. FLAN T5's error-correcting capabilities \cite{nanayakkara2022clinical,nguyen2020neural} aid in reconstructing sentence fragments after words or phrases with elevated levels of concern have been removed.
The user interface is built with standard HTML, CSS, and JavaScript. Both components are open source and on GitHub\footnote{\url{https://github.com/dimitristaufer/Semi-Automated-Text-Sanitization}}.

\subsection{Anonymization Operations for Words and Phrases}
\label{sec_impl_1}

\subsubsection{Generalization} \label{details_generalization}

\textcolor{mediumconcerndarker}{\texttt{\textbf{Common nouns}}} undergo generalization by first retrieving their synsets and hypernyms from WordNet, followed by calculating the cosine similarity of their sentence embeddings with those of the hypernyms. 
This calculation ranks the hypernyms by semantic similarity to the original word, enabling the selection of the most suitable replacement. By default, we select the closest hypernym. \textcolor{mediumconcerndarker}{\texttt{\textbf{Other proper nouns}}} are generalized as follows: We first query Wikipedia to identify the term, using the \textit{all-mpnet-base-v2} sentence transformer to disambiguate its meaning through cosine similarity. Next, we find the most relevant Wikidata QID and its associated hierarchy. We then flatten these relationships and replace the entity with the next higher-level term in the hierarchy.

\subsubsection{Perturbation} \label{details_perturbation}

We add randomness to \textcolor{mediumconcerndarker}{\texttt{\textbf{modifiers}}} and \textcolor{mediumconcerndarker}{\texttt{\textbf{named entities}}} through LLM-based paraphrasing, specifically, by using the FLAN-T5 language model, which we fine-tuned for paraphrase generation (Section \ref{sec_impl_2}). To achieve perturbation%
\footnote{The strategies ``suppression'' and ``generalization'' 
are straightforward adaptations of the classical methods for structured data. Perturbation ``replaces original values with new ones by interchanging, adding noise or creating synthetic data'' \cite{perturbationref}.  Interchanging would create ungrammatical texts, and noise can only be added to certain data. We, therefore, generate synthetic data via LLM-Rephrasing, disallowing the highly specific words / terms and their synonyms while producing a new but grammatical text.}, we give the tokens in question and their synonyms and inflections zero weight during next token prediction. This forces the model to either use a less probable word (controlled by the \textit{temperature} hyperparameter) or rephrase the sentence to omit the token. Using a LLM for paraphrase generation has the added benefit that it mends fragmented sentences caused by token suppression and yields a neutral writing style, adjustable through the \textit{no\_repeat\_ngram\_size} hyperparameter.

\subsubsection{Suppression} \label{details_suppression}

\textcolor{highconcerndarker}{\texttt{\textbf{Common nouns}}} and \textcolor{highconcerndarker}{\texttt{\textbf{other proper nouns}}} are suppressed by removing the longest phrase containing them with the \textit{constituent\_treelib} library. Sentences with just one noun or proper noun are entirely removed. Otherwise, the longest phrase, be it a main clause, verb phrase, prepositional phrase, or noun phrase, is identified, removed, and replaced with an empty string.
\textcolor{highconcerndarker}{\texttt{\textbf{Modifiers}}} are removed (e.g., \enquote{He is their \underline{principal} dancer} $\rightarrow$ \enquote{He is their · dancer}). \texttt{\textcolor{mediumconcerndarker}{\textbf{Pro}}\textcolor{highconcerndarker}{\textbf{nouns}}} are replaced with the static string \enquote{somebody}. For example, \enquote{\underline{His} apple} $\rightarrow$ \enquote{\underline{Somebody} apple} (after replacement) $\rightarrow$ \enquote{Somebody's apple} (after paraphrase generation).
\textcolor{highconcerndarker}{\texttt{\textbf{Named entities}}} are replaced with static phrases based on their type. For example, \enquote{\underline{John Smith} sent her \underline{2 Million Euros} from his account in \underline{Switzerland}} $\rightarrow$ \enquote{\underline{certain person} sent somebody \underline{certain money} from somebody account in \underline{certain location}} (after suppressing pronouns and named entities) $\rightarrow$ \enquote{A certain individual sent a specific amount of money to whoever's account in some particular place} (after paraphrase generation).

\subsection{Paraphrase Generation}
\label{sec_impl_2}

We fine-tuned two variants of the FLAN T5 language models, FLAN T5$_{\text{Base}}$ and FLAN T5$_{\text{XL}}$, using the \enquote{chatgpt-paraphrases} dataset, which uniquely combines three large paraphrasing datasets for varied topics and sentence types. It includes question paraphrasing from the \enquote{Quora Question Pairs} dataset, context-based paraphrasing from \enquote{SQuAD2.0}, and summarization-based paraphrases from the \enquote{CNN-DailyMail News Text Summarization} dataset. Furthermore, it was enriched with five diverse paraphrase variants for each sentence pair generated by the \textit{gpt-3.5-turbo} model, resulting in 6.3 million unique pairs. This diversity enhances our model’s paraphrasing capabilities and reduces overfitting.

For training, we employed Parameter-Efficient Fine-Tuning (\textit{PEFT}) using \textit{LoRA} (Low-Rank Adaptation), which adapts the model to new data without the need for complete retraining. We quantized the model weights to enhance memory efficiency using \textit{bitsandbytes}. We trained FLAN T5$_{\text{Base}}$ on a NVIDIA A10G Tensor Core GPU for one epoch (35.63 hours) on 1 mio. 
paraphrase pairs, using an initial learning rate of 1e-3. After one epoch, we achieved a minimum Cross Entropy loss of 1.195. FLAN T5$_{\text{XL}}$ was trained for one epoch (22.38 hours) on 100,000 pairs and achieved 0.88.

For inference, we configure \texttt{max\_length} to 512 tokens to cap the output at T5's tokenization limit. \texttt{do\_sample} is set to \texttt{True}, allowing for randomized token selection from the model's probability distribution, enhancing the variety of paraphrasing. Additionally, parameters like \texttt{temperature}, \texttt{no\_repeat\_ngram\_size}, and \texttt{length\_penalty} are adjustable via the user interface, providing control over randomness, repetition avoidance, and text length.

\subsection{User Interface}
\label{sec_impl_3}

\begin{figure}[h]
  \centering
  \includegraphics[width=\linewidth]{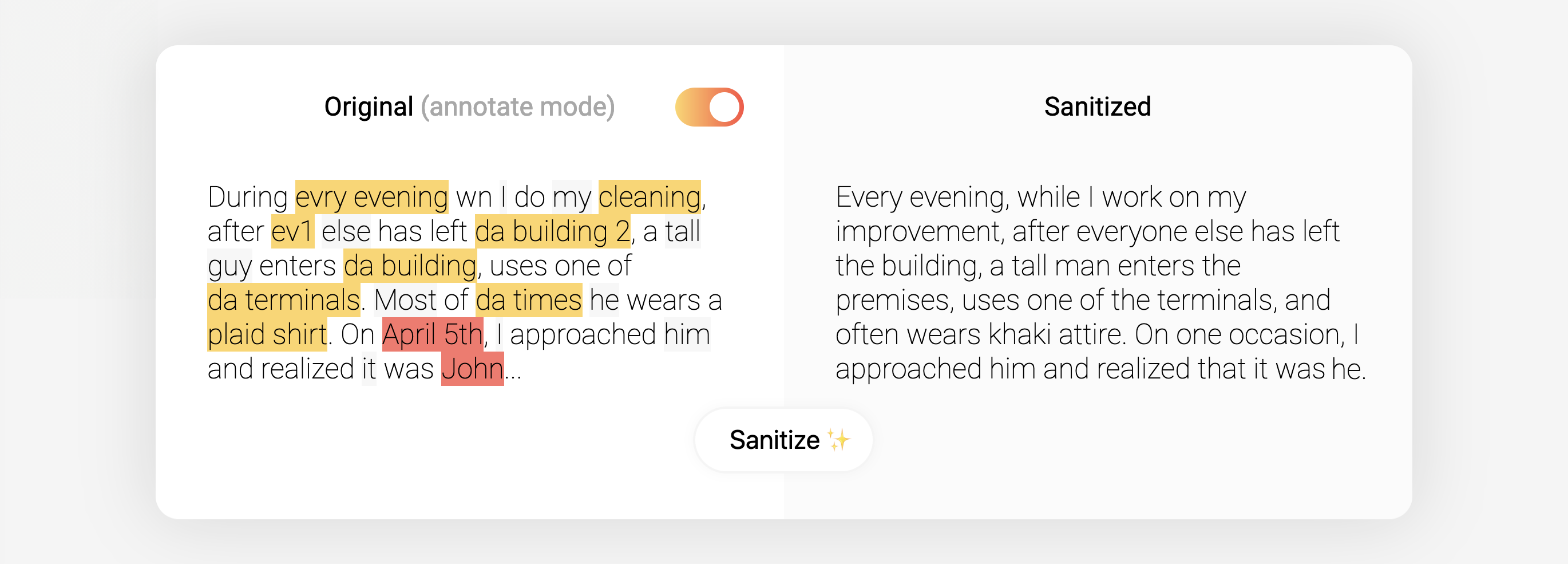}
  \caption{The UI's main window showing the input text (left) and the sanitized text (right). We made up the input and converted it to \enquote{Internet Slang} (\url{https://www.noslang.com/reverse}) to showcase how an extremely obvious writing style is neutralized.}
  \Description{The UI's main window showing the input text (left) and the sanitized text (right). We made up the input and converted it to \enquote{Internet Slang} (\url{https://www.noslang.com/reverse}) to showcase how an extremely obvious writing style is neutralized.}
  \label{fig:user_interface}
\end{figure}

Our web-based user interface communicates with the sanitization pipeline via \textit{Flask} endpoints. It visualizes token LvCs (gray, yellow, red), allows dynamic adjustments of these levels, and starts the sanitization process. Moreover, a responsive side menu allows users to select the model size and tune hyperparameters for paraphrasing. The main window (Figure \ref{fig:user_interface}) shows the original and the sanitized texts, with options for editing and annotating.

\section{Evaluation}
\label{sec_eval}

We evaluate \toolname\ quantitatively (Sections \ref{sec_imdb} and \ref{echr_cases}) and demonstrate its workings and usefulness with an example from a real-world whistleblower testimony (Section \ref{sec_hunter}). They complement each other in that the first focuses on identification via writing style and the second two on identification via content.

\subsection{Re-Identification Through Writing Style: IMDb62 Movie Reviews Dataset}
\label{sec_imdb}

\subsubsection{Evaluation metrics}
\label{sec_evaluation_design}

The large unknowns of context knowledge imply that evaluations cannot rely on straightforward measurement methods for $A_{id}$ and $E_{dt}$. We, therefore, work with the following proxies.
\begin{description}

\item[Text-surface similarities] To understand the effect of language model size and hyperparameter settings on lexical and syntactic variations from original texts, we utilize two ROUGE scores: ROUGE-L (Longest Common Subsequence) to determine to which extent the overall structure and sequence of information in the text changes. And ROUGE-S (Skip-Bigram) to measure word pair changes and changes in phrasings.
\item[Risk] 
Without further assumptions about the (real-world case-specific) background knowledge, it is impossible to exactly quantify the ultimate risk of re-identification (see Section \ref{sec_identifier_types}). We therefore only measure the part of $A_{id}$ where (a) the context knowledge is more easily circumscribed (texts from the same author) and (b) benchmarks are likely to generalize across case studies: the risk of re-identification based on stylometric features, measured as authorship attribution accuracy (AAA).
\item[Utility] It is also to be expected that the rewriting reduces $E_{dt}$, yet again it is impossible to exactly determine (without real-world case-specific background knowledge and legal assessment) whether the detail supplied is sufficient to allow for legal follow-up of the report or even only to create alarm that could then be followed up. We, therefore, measure $E_{dt}$ utility through two proxies:
a semantic similarity measure and a sentiment classifier. To estimate semantic similarity (\textbf{SSim}), we calculate the cosine similarity of both texts' sentence embeddings using the SentenceTransformer\footnote{all-mpnet-base-v2} Python framework. To determine the absolute sentiment score difference (\textbf{SSD}), we classify the texts' sentiment using an off-the-shelf BERT-based classifier\footnote{bert-base-multilingual-uncased} from Hugging Face Hub. 
\end{description}

All measures are normalized to take on values between 0 and 1, and although the absolute values of the scores between these endpoints (except for authorship attribution) cannot be interpreted directly, the comparison of relative orders and changes will give us a first indication of the impacts of different rewriting strategies on $A_{id}$ and $E_{dt}$.

\subsubsection{Data, language models, and settings}

We investigate protection against authorship attribution attacks with the popular IMDb62 movie reviews dataset \cite{seroussi2014authorship}, which contains 62,000 movie reviews by 62 distinct authors.
We assess AAA using the \enquote{Valla} software package \cite{tyo2022state}, specifically its two most effective models: one based on character \textit{n}-grams and the other on BERT. This approach covers both ends of the the authorship attribution spectrum \cite{altakrori2021topic}, from low-level, largely topic-independent character \textit{n}-grams to the context-rich features of the pre-trained BERT model.

The evaluation was conducted on AWS EC2 \enquote{g4dn.xlarge} instances with NVIDIA T4 GPUs. We processed 130 movie reviews for each of the 62 authors across twelve FLAN T5 configurations, totaling 96,720 texts with character counts spanning from 184 to 5248. Each review was sanitized with its textual elements assigned their default LvCs (see \ref{automatic_LvC_estimation}).

Both model sizes, \enquote{Base} (250M parameters) and \enquote{XL} (3B parameters) were tested with temperature values T of 0.2, 0.5, and 0.8, as well as with \texttt{no\_repeat\_ngram\_size} (NRNgS) set to 0 or 2. The former, \textit{temperature}, controls the randomness of the next-word predictions by scaling the logits before applying softmax, which makes the predictions more or less deterministic. For our scenario, this causes smaller or greater perturbation of the original text's meaning. The latter, NRNgS, disallows n consecutive tokens to be repeated in the generated text, which for our scenario means deviating more or less from the original writing style.

The Risk-utility trade-offs of all configurations are compared to three baselines: $Baseline_{1}$ is the original text. In $Baseline_{2}$, similar to state-of-the-art related work \cite{kleinberg2022textwash, papadopoulou2022neural}, we only redact named entities by replacing them with placeholders, such as \enquote{[PERSON]} and do not utilize our language model. Similarly, in $Baseline_{3}$ we only remove named entities but rephrase the texts using our best-performing model configuration regarding AA protection.

\subsubsection{Results}

The n-gram-based and BERT-based \enquote{Valla} classifiers achieved AAA baselines of \texttt{98.81}\% and \texttt{98.80}\%, respectively. As expected, the AAA and text-surface similarities varied significantly depending on the model configuration. The XL-model generated texts with much smaller ROUGE-L and ROUGE-S scores, i.e. more lexical and syntactic deviation from the original texts. Using $NRNgS = 2$ slightly decreased AAA in all configurations while not significantly affecting semantic similarity, which is why we use this for all the following results.

\begin{figure}[H]
  \centering
  \subfloat[Risk-utility trade-off between AAA and SSim.]{
    \includegraphics[width=0.47\textwidth]{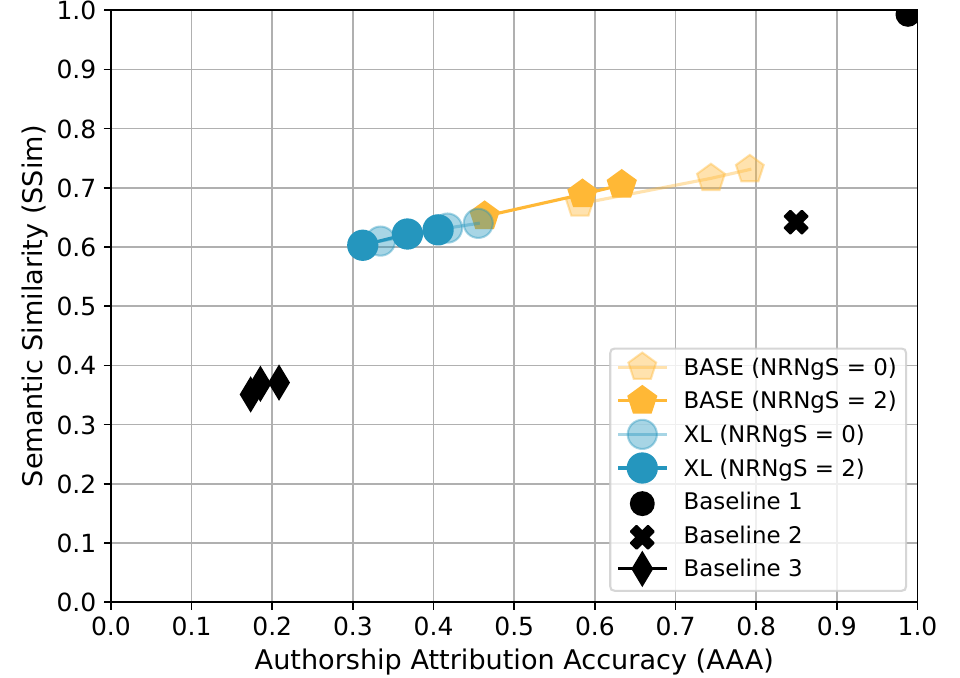}
    \label{fig:a}
  }\vspace{-3pt}
  \subfloat[Risk-utility trade-off between AAA and SSD.]{
    \includegraphics[width=0.47\textwidth]{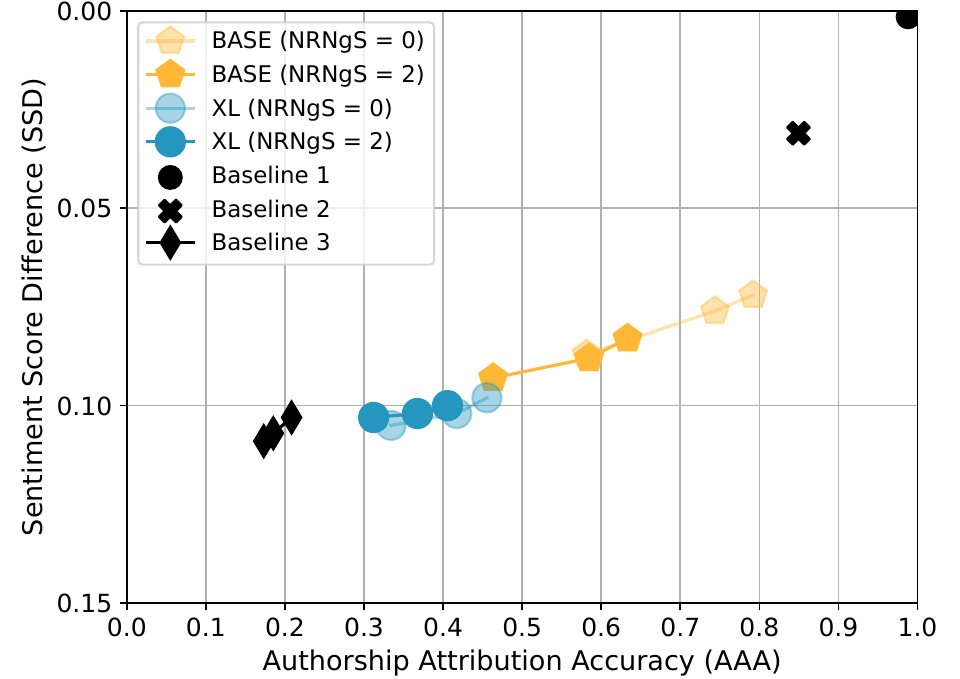}
    \label{fig:b}
  }\vspace{-3pt}
  \caption{Risk-utility trade-offs.}
  \Description{Risk-utility trade-offs.}
  \label{fig:quant_eval_fig_2}
\end{figure}

Figure \ref{fig:quant_eval_fig_2} (a) shows the risk-utility trade-off between AAA and SSim. ``Top-left'' (0,1) would be the - fictitious - best result. For each model configuration, increasing $T$ caused AAA to drop but also decreased utility by $\sim8\%/4\%$ (BASE/XL) for SSim and $\sim$\texttt{12}\%/\texttt{3}\% (BASE/XL) for SSD. The figure shows that the investigated settings create a trade-off curve, with XL ($T=0.8$, $NRNgS=2$) allowing for a large reduction in AAA (to \texttt{31.22}\%, as opposed to the original text $baseline_{1}$ of 98.81\%), while BASE ($T=0.2$, $NRNgS=0$) retains the most SSim (0.731, as opposed to the original texts, which have $SSim=1$ to themselves).

Figure \ref{fig:quant_eval_fig_2} (b) shows the risk-utility trade-off between AAA and SSD (the plot shows 1-SSD to retain ``top left'' as the optimal point). The results mirror those of AAA-SSim, except for $baseline_{2}$: because only named entities (not considered sentiment-carrying) are removed, the sentiment score changes only minimally.

\subsubsection{Discussion}

In summary, all our models offer a good compromise between baselines representing state-of-the-art approaches. They have lower risk and higher or comparable utility compared to $baseline_{3}$, where only named entities are removed. This indicates the effectiveness of LLM-based rephrasing in authorship attribution. $Baseline_{2}$, which involves suppressing named entities and rephrasing, shows the lowest risk due to limited content left for the LLM to reconstruct, resulting in mostly short, arbitrary sentences, as reflected by low SSim scores.

\subsection{Re-Identification Through Content: European Court of Human Rights Cases}
\label{echr_cases}

Pilán et al.’s \cite{pilan2022text} Text Anonymization Benchmark (TAB) includes a corpus of 1,268 English-language court cases from the European Court of Human Rights, in which directly- and quasi-identifying nominal and adjectival phrases were manually annotated. It solves several issues that previous datasets have, such as being \enquote{pseudo-anonymized}, including only few categories of named entities, not differentiating between identifier types, containing only famous individuals, or being small. TAB's annotation is focused on protecting the identity of the plaintiff (also referred to as \enquote{applicant}).

\subsubsection{Evaluation Metrics}

TAB introduces two metrics, entity-level recall ($ER_{di/qi}$) to measure privacy protection and token-level\- weighted precision ($WP_{di+qi}$) for utility preservation. Entity-level means that an entity is only considered safely removed if all of its mentions are. $WP_{di+qi}$ uses BERT to determine the information content of a token \textit{t} by estimating the probability of \textit{t} being predicted at position \textit{i}. Thus, precision is low if many \textit{t} with high information content are removed. Both metrics use micro-averaging over all annotators to account for multiple valid annotations. Because our tool automatically rephrases the anonymized texts, we make two changes. First, since we cannot reliably measure $WP_{di+qi}$, we fall back to our previously introduced proxies for measuring $E_{dt}$ utility. Secondly, we 
categorize newly introduced entities from LLM hallucination that may change the meaning of the sanitized text.

The legal texts, which must prefer direct and commonly-known identifiers, are likely to present none or far fewer of the background-knowledge-specific re-identification challenges of our domain. Thus, again the metrics used here should be regarded as proxies.


\begin{description}

\item[Risk] We measure $A_{id}$ using $ER_{di/qi}$ and count slightly rephrased names of entities as \enquote{not removed} using the Levenshtein distance. For example, rephrasing \enquote{USA} as \enquote{U.S.A} has the same influence on $ER_{di/qi}$ as failing to remove \enquote{USA}.
\item[Utility] We estimate $E_{dt}$ through \textit{SSim}. In addition, we determine all entities in the sanitized text that are not in the original text (again using the Levenshtein distance). We categorize them into (1) \textit{rephrased harmful entities} (semantically identical to at least one entity that should have been masked), (2) \textit{rephrased harmless entities}, and (3) \textit{newly introduced entities}. We measure semantic similarity by calculating the cosine similarity of each named entity phrase's sentence embedding to those in the original text.

\end{description}

\subsubsection{Data, language models, and settings}

The TAB corpus comprises the first two sections (introduction and statement of facts) of each court case. For our evaluation, we use the \textit{test} split which contains 127 cases of which each has, on average, 2174 characters (356 words) and 13.62 annotated phrases. We perform all experiments using the \enquote{XL} (3B parameter) model with temperature values T of 0.2, 0.5, and 0.8, as well as with \textit{NRNgS} set to 2. 

\subsubsection{Results and Discussion}

$ER_{di/qi}$ and SSim vary slightly, but not significantly for different T values. For \textit{T = 0.2}, we get an entity-level recall on quasi-identifiers ($ER_{qi}$) of 0.93, which is slightly better than Pilán et al.’s \cite{pilan2022text} best performing model trained directly on the TAB corpus (0.92). However, our result for direct identifiers $ER_{di}$ is 0.53, while theirs achieves 1.0, i.e. does not miss a single high-risk entity. Closer inspection reveals that our low results for direct identifiers come mainly from (i) the SpaCy NER failing to detect the entity type CODE (e.g. \enquote{10424/05}) and (ii) the LLM re-introducing names of named entities that are spelled slightly differently (e.g. \enquote{Mr Abdisamad Adow Sufi} instead of \enquote{Mr Abdisamad Adow Sufy}). 

Regarding utility, all three model configurations achieve similar SSim scores ranging from 0.67 (T = 0.8) to 0.69 (T = 0.2). These results fall into the same range achieved using the IMDb62 movie reviews dataset. However, in addition to re-introducing entities that should have been masked, we found that, on average, the LLM introduces 5.24 new entities (28.49\%) per court case. While some of these, depending on the context, can be considered harmless noise (e.g. \enquote{European Supreme Tribunal}), manual inspection revealed that many change the meaning and legitimacy of the sanitized texts. For example, 4.7\% contain names of people that do not appear in the original text, 43.3\% contain new article numbers, 20.5\% contain new dates, and 11.8\% include names of potentially unrelated countries.

The frequency of such hallucinations could also be a consequence of the specific text genre of court cases, and future work should examine to what extent this also occurs in whistleblower testimonies and how it affects the manual post-processing over the generated text that is previewed in our semi-automated tool.

\subsection{Re-Identification Through Content: Whistleblower Testimony Excerpts}
\label{sec_hunter}

We further investigated \toolname's rewritings of two excerpts (Tables \ref{tab:qualitative_comparison_1}, \ref{tab:qualitative_comparison_2}) from a whistleblower's hearing in the Hunter Biden tax evasion case, as released by the United States House Committee on Ways and Means.\footnote{\url{https://waysandmeans.house.gov/?p=39854458} [Accessed 29-April-2024], \enquote{\#2}} This qualitative view on our results provides for a detailed understanding of which identifiers were rewritten and how.\footnote{To answer these questions, it is immaterial whether the text sample describes a concrete act of wrongdoing (as in our fictitious Ex. 1) or not (as here).}

\subsubsection{Approach}

First, we compiled the essential $E_{dt}$ upon which we based our analysis on. Next, we assessed the textual features in both excerpts to enhance \toolname's automatic Level of Concern (LvC) estimations, aiming for the lowest author identifiability ($A_{id}$). Finally, we input these annotations into the user interface to produce the rewritings.

\subsubsection{$E_{dt}$ and $A_{id}$}

Based on the information from the original texts in tables \ref{tab:qualitative_comparison_1} and \ref{tab:qualitative_comparison_2} alone, we define $E_{dt}$ as follows, with \( E_{dt1} \), \( E_{dt2} \) being a subset of excerpt 1 and  \( E_{dt3} \) a subset of excerpt 2.
{
  \centering
  \includegraphics[width=\linewidth, trim={0cm 0.02cm 0cm 0.02cm}, clip]{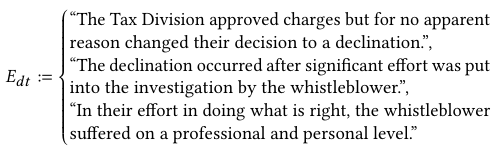}
  \par 
}

In $exc1$ (Table \ref{tab:qualitative_comparison_1}), we classified \enquote{joining the case} (first-person indexical) and implications of a nation-wide investigation as highly concerning. Additionally, we marked all \enquote{case} mentions as highly concerning to evaluate consistent suppression. \enquote{DOJ Tax}, being a stylometric identifier because it is 
no official abbreviation, received a medium LvC, and \enquote{thousands of hours} was similarly categorized, potentially indicating the authors role as lead in the case.

In $exc2$ (Table \ref{tab:qualitative_comparison_2}), we classified the lexical identifier \enquote{2018}, which could be cross-referenced relatively easily, as well as all descriptive identifiers concerning the author's sexual orientation and outing as highly concerning. Furthermore, emotional descriptors (\enquote{sleep, vacations, gray hairs, et cetera}) are given medium LvC, similar to references of case investment (\enquote{thousands of hours} and \enquote{95 percent}), mirroring the approach from $exc1$.

\subsubsection{Results and Discussion}

$Exc1_{sanitized}$ retains \( E_{dt2} \), but not \( E_{dt1} \), as ``DOJ Tax'' is replaced with ``proper noun'' due to the non-existence of a corresponding entity in Wikidata. Consequently, it defaults to the token's POS tag. For \( A_{id} \), all identified risks were addressed (e.g., ``considerable time'' replaces ``thousands of hours.''). However, the generalization of ``case'' led to inconsistent terms like ``matter'', ``situation'', and ``issue'' due to the \( NRNgS = 2 \) setting. This is beneficial for reducing authorship attribution accuracy but may confuse readers not familiar with the original context.

$Exc2_{sanitized}$ maintains parts of \( E_{dt3} \), though terms like ``X amount of time'' and ``Y amount of the investigation'' add little value due to their lack of specificity. Notably, ``amount o of'' represents a rare LLM-induced spelling error, underscoring the need for human editing for real-world use. The emotional state's broad generalization to ``physical health, leisure, grey body covering'' is odd and less suitable than a singular term would be. Despite this, $Exc2_{sanitized}$ effectively minimizes \( A_{id} \) by addressing all other identified risks.

\newcommand{\mediumconcerndarker}[1]{\textbf{\textcolor{mediumconcerndarker}{#1}}} 
\newcommand{\highconcerndarker}[1]{\textbf{\textcolor{highconcerndarker}{#1}}} 

\newcommand{\topstrut}[1]{\rule{0pt}{#1}}     
\newcommand{\bottomstrut}[1]{\rule[-#1]{0pt}{0pt}}  

\noindent
\begin{table}[h]
\caption{LvC-annotated whistleblower testimony \( exc_1 \) (excerpt 1) with identifiers (top) and \( exc1_{sanitized} \) (bottom).}
\Description{LvC-annotated whistleblower testimony \( exc_1 \) with identifiers (top) and \( exc1_{sanitized} \) (bottom).}
\centering
\renewcommand{\arraystretch}{1.0}
\begin{tabular}{| p{\linewidth} |}
\hline
\topstrut{10pt}
\textbf{Original:} \enquote{\highconcerndarker{Prior} to joining the \highconcerndarker{case}, \mediumconcerndarker{DOJ Tax} had approved tax charges for the \highconcerndarker{case} and the \highconcerndarker{case} was in the process of progressing towards indictment [...] After working \mediumconcerndarker{thousands of hours} on that captive \highconcerndarker{case}, poring over evidence, interviewing \highconcerndarker{witnesses} all over the \highconcerndarker{U.S.}, the decision was made by \mediumconcerndarker{DOJ Tax} to change the approval to a declination and not charge the \highconcerndarker{case}.} \\
\textbf{Lexical IDs:} DOJ Tax; U.S. \\
\textbf{Indexical IDs:} [implicit: me] joining the case (first person) \\
\textbf{Descriptive IDs:} interviewing witnesses all over the U.S. (nation-wide investigation); thousands of hours (author involvement) \bottomstrut{5pt} \\
\hline
\topstrut{10pt}
\textbf{Sanitized:} \enquote{The proper noun had approved tax charges for the matter and the situation was moving towards indictment, but after spending considerable time on that captive matter, poring over evidence, the decision was made by proper noun to defer the approval and not charge the issue.} \bottomstrut{5pt} \\
\hline
\end{tabular}
\label{tab:qualitative_comparison_1}
\end{table}

\noindent
\begin{table}[h]
\caption{LvC-annotated whistleblower testimony \( exc2 \) (excerpt 2) with identifiers (top) and \( exc2_{sanitized} \) (bottom).}
\Description{LvC-annotated whistleblower testimony \( exc2 \) with identifiers (top) and \( exc2_{sanitized} \) (bottom).}
\centering
\renewcommand{\arraystretch}{1.0}
\begin{tabular}{| p{\linewidth} |}
\hline
\topstrut{10pt}
\textbf{Original:} \enquote{I had opened this investigation in \highconcerndarker{2018}, have spent \mediumconcerndarker{thousands of hours} on the case, worked to complete \mediumconcerndarker{95 percent} of the investigation, have sacrificed \mediumconcerndarker{sleep}, \mediumconcerndarker{vacations}, \mediumconcerndarker{gray} \mediumconcerndarker{hairs}, \mediumconcerndarker{et cetera}. \highconcerndarker{My} \highconcerndarker{husband} and I, in identifying \highconcerndarker{me} as the \highconcerndarker{case agent}, were \highconcerndarker{both} \highconcerndarker{publicly} outed and ridiculed on social media due to \highconcerndarker{our} \highconcerndarker{sexual} \highconcerndarker{orientation}.} \\
\textbf{Lexical IDs:} 2018; thousands of hours; 95 percent \\
\textbf{Indexical IDs:} me as the case agent (role of author); My husband (author's marital status) \\
\textbf{Descriptive IDs:} I had opened this investigation in 2018 (can be cross-referenced); My husband and I + publicly outed and ridiculed [...] due to our sexual orientation (author's sexual orientation and public event); sacrificed sleep, [...], gray hairs (emotional state) \bottomstrut{5pt} \\
\hline
\topstrut{10pt}
\textbf{Sanitized:} \enquote{I had opened this investigation on a certain date, had spent X amount of time on the case, worked to complete Y amount of the investigation, sacrificing my physical health, leisure, grey body covering, etc.} \bottomstrut{5pt} \\
\hline
\end{tabular}
\label{tab:qualitative_comparison_2}
\end{table}

\section{Conclusions, Limitations and Future Work}
\label{sec_summary}

We evaluated our \toolname's effectiveness using ECHR court cases and excerpts from a real-world whistleblower testimony and measured the protection against authorship attribution attacks and information loss statistically using the popular IMDb62 movie reviews dataset. Our method can significantly reduce authorship attribution accuracy from 98.81\% to 31.22\%, while preserving up to 73.1\% of the original content's semantics, as measured by the established cosine similarity sentence embeddings.
Our qualitative analysis revealed that minor wording changes significantly impact $A_{id}$ and $E_{dt}$, and highlighted \toolname's strengths in reducing $A_{id}$ through generalization, perturbation, and suppression.

Our tool's usefulness in real-world whistleblowing scenarios remains to be tested, particularly with human users. Challenges arise from the possibility of the tool introducing unrelated entities through model hallucination and its limitations in addressing complex syntactic structures and co-references. Still, our LLM-based approach has proved to be promising in matters of counteracting the limitations of state-of the art approaches. The fine-tuned model effectively reduces authorship attribution and improves text coherence – two of the main shortcomings of previous works. At the same time, it introduces novel challenges, such as limited control over the accuracy and consistency of the rephrased content.



Future work will focus on refining \toolname\ through evaluations involving human participants and domain experts. Given the crucial importance of context knowledge for re-identification risks and the challenges in identifying all textual features that contribute to re-identification, future work will also pay increasing attention to enhancing anonymization awareness. This would not only apply to the whistleblowing use case, but extend to the protection of free speech in other areas too, including journalism, political activism, and social media.

We envision an interactive awareness tool as a more dynamic alternative to conventional static writing guides on whistleblowing platforms. This tool would incorporate insights from our research as well as insights from practitioners, aiming to educate users about subtle textual nuances that could pose re-identification risks, thereby creating a deeper understanding and more effective use of anonymization practices in high-risk disclosures. 
At the same time, we need to draw on practitioners' and legal experts' knowledge to better understand what textual changes are detrimental (or conducive) to utility and incorporate these insights into the guidance provided by the awareness tool.


\section{Ethical considerations, researchers positionality, and possible adverse impacts}
\label{sec_ethics}

In the following paragraphs, we discuss five key challenges, interweaving a potential adverse impacts statement, an ethical considerations statement (what we have done or can do), and positionalities. 

We are computer scientists (some of us with a background also in social and legal sciences) 
who have programming expertise (instrumental for mitigating challenges C1--C4), understanding of data protection law (C1), research expertise in bias and fairness, including methods for risk mitigation when working with LLMs (C2), and collaborators with human-subjects studies expertise (C3). None of us has been a whistleblower. We outline below how future collaborators and/or deployers with other positionalities can contribute relevant complementary expertise on C1--C5.

\vspace{0.2cm}

\noindent{\em C1 -- Data Protection:}
Our tool does not collect or store any user data. Original as well as re-written texts are discarded after each run, and they are not used to train the model further. Our tool does not require an internet connection beyond the initial downloading of pre-trained language models and optional queries to Wikidata servers. 
While querying Wikidata enhances the efficacy of our tool by enabling the generalization of certain words, users
should be aware that these queries might expose confidential information
to external servers. To mitigate this risk, our implementation remains functional when offline, albeit with slightly reduced efficacy due to the lack of real-time Wikidata look-ups.
In a real-life deployment, technical and organizational measures would need to be implemented in order to safeguard the confidential personal or organizational data that remain in the reports; this will also require security and legal expertise.  

\noindent{\em C2 -- Bias and (Un-)fairness:}
Our tool may inadvertently introduce or perpetuate biases present in the training data. FLAN T5 was trained on C4, which is generated from the April 2019 Common Crawl dataset. \citet{dodge2021documenting} discovered that C4 has a \enquote{negative sentiment bias against Arab identities} and excludes \enquote{documents associated with Black and Hispanic authors} as well as documents \enquote{mentioning sexual orientations} [p. 8] by its blocklist filter. Therefore, similar to other pre-trained models \cite{mao2022biases}, FLAN T5 is \enquote{potentially vulnerable to generating equivalently inappropriate content or replicating inherent biases} \cite[p. 52]{chung2022scaling}.
This may bias our level of concern measures.
For example, certain names, professions, or locations may be classified as \enquote{medium concerning} or \enquote{highly concerning} more often because they are considered \enquote{surprising}, which may unfairly impact the narratives involving them. Future work should, therefore, include evaluating and mitigating these biases and possibly experiments with other datasets and pre-trained models.

\noindent{\em C3 -- Over-Reliance and Retaliation:}
The results of our quantitative evaluation are promising, but an extensive qualitative evaluation is necessary to determine whether our approach translates to real-world situations. Therefore, users of \toolname\ must remain aware of its potential to alter the original intent of their text significantly and, depending on the context, possibly offer limited protection against retaliation. Over-reliance on \toolname\ may lead to a false sense of security, resulting in increased vulnerability to retaliation.
We intend to assess the extent of this form of
automation bias \cite{automationbias} in a subsequent user study,  discuss with people who are working in the field (e.g., whistleblower protection activists) how to best reduce it, and also evaluate these future mitigation measures.


\noindent{\em C4 -- Resource consumption:}
Training LLMs is resource-intensive. By re-using the existing model and enlisting distilled LLM learning, this impact could be reduced in future work. 

\noindent{\em
C5
-- Tool Misuse:}
Even though \toolname\ aims to mitigate the risk of whistleblower re-identification, malicious actors might misuse \toolname\ for obfuscating dangerous information or illegally converting copyrighted material. By providing our source code and fine-tuned models publicly, we open avenues for ethical use and misuse alike. Therefore, we emphasize that our sole aim in developing \toolname\ is to facilitate legal, ethical whistleblowing.
Future refinements and real-world evaluations will require collaboration with legal and social experts to better understand the practical implications and potential misuse scenarios.


\begin{acks}

We acknowledge funding from the German Federal Ministry of Education and Research (BMBF) – Nr 16DII134.

This publication has been partially supported by the EXDIGIT (Excellence in Digital Sciences and Interdisciplinary Technologies) project, funded by Land Salzburg under grant number 20204-WISS/\allowbreak263/6-6022.

\end{acks}

\bibliographystyle{ACM-Reference-Format}
\bibliography{references}

\end{document}